\documentclass[aps,prb,preprint,showpacs,groupedaddress]{revtex4}
\usepackage[dvips]{graphicx}
\begin{document}
\title{Kinetic arrest of the first order ferromagnetic to antiferromagnetic transition in Ce(Fe$_{0.96}$Ru$_{0.04}$)$_2$ : formation of a magnetic-glass}
\author{M. K. Chattopadhyay$^1$}
\author{S. B. Roy$^1$}
\email {sbroy@cat.ernet.in} 
\author{P. Chaddah$^{1,2}$}
\affiliation{$^1$ Magnetic and Superconducting Materials Section, Centre for Advanced Technology, Indore 452013, India.}
\affiliation{$^2$UGC-DAE Consortium for Scientific Research,
University Campus, Khandwa Road,
Indore 452017, India.}
\date{\today}
\begin{abstract}
We present results of dc magnetization and magnetic relaxation study showing the kinetic arrest of a first order ferromagnetic to antiferromagnetic transition in Ce(Fe$_{0.96}$Ru$_{0.04}$)$_2$. This leads to the formation of a non-ergodic glass-like magnetic state. The onset of the magnetic-glass transformation is tracked through the slowing down of the magnetization dynamics. This glassy state is formed with the assistance of an external magnetic field and this is distinctly different from the well known 'spin-glass' state. 
\end{abstract} 
\maketitle
Liquids freeze into crystalline solids via a first order phase transition. However, some liquids called 'glass formers' experience a viscous retardation of nucleation and crystallization in their supercooled state. In the experimental time scale the supercooled liquid ceases to be ergodic and it enters a glassy state. It is a great challenge to fully understand the nature of glass transition\cite{1,2}, and perhaps it is the deepest and most important unsolved problem in condensed matter physics\cite{3}. In this respect one of the most provocative aspects concerns the slowing down of the dynamics on decreasing the temperature of the melt\cite{1,4,5}. We show here the same phenomena in the formation of a magnetic field assisted glass-like magnetic state in Ce(Fe$_{0.96}$Ru$_{0.04}$)$_2$. In striking contrast with 'spin-glass'\cite{6} this magnetic state arises out of a kinetically arrested first order ferromagnetic to antiferromagnetic phase transition. Experimentally, the exploration of pressure-temperature phase space in the context of glass formation has remained a difficult proposition due to the obvious problems with pressure studies\cite{7}.In this regard such a magnetic medium will provide an excellent context to study the properties of glass in magnetic field-temperature phase space.

CeFe$_2$ is a cubic Laves phase ferromagnet (with T$_{Curie} \approx$230K)\cite{8} where small substitution ($<$10 \%) of selected  elements such as Co, Al, Ru, Ir, Os and Re can induce a low temperature antiferromagnetic state\cite{9}. This ferromagnetic (FM) to antiferromagnetic (AFM) transition in doped-CeFe$_2$ alloys has been a subject for many studies \cite{9,10,11} and the first order nature of the transition is well established through various bulk properties measurements\cite{12,13,14}, neutron scattering\cite{10}, microcalorimetry\cite{15} and micro-Hall probe imaging\cite{16}.  We use here a 4\%Ru-doped CeFe$_2$ pseudobinary alloy to show that the first order FM to AFM transition process is arrested when the applied field is beyond a certain critical value, and the lowest temperature state becomes non-ergodic in nature. The observed phenomenon has the characteristics of a glass transition. It should be noted here that the behaviour typical of glass formation is not necessarily restricted to materials that are positionally disordered\cite{4}. Such phenomenon has been observed for apparently crystalline materials composed of asymmetric molecules\cite{4}. Even for the conventional glasses, other than the general definition that 'glass is a noncrystalline solid material which yields broad nearly featureless diffraction pattern', there exists another widely acceptable picture of glass as a liquid where the atomic or molecular motions are arrested. Within this latter dynamical framework 'glass is time held still'\cite{4}.

The Ce(Fe$_{0.96}$Ru$_{0.04}$)$_2$ polycrystalline alloy used in the present study was prepared by argon-arc melting. The details of the sample preparation and characterization  have been described earlier\cite{9,10,16}.  Neutron diffraction studies of the same sample revealed a discontinuous change of the unit cell volume at the FM-AFM transition,  confirming that it is first order\cite{10}. Bulk magnetization measurements were made with a commercial SQUID magnetometer (Quantum Design-MPMS5). Three different experimental protocols, zero field cooled(ZFC), field cooled cooling (FCC) and field cooled warming (FCW)are used for magnetization measurements. In ZFC mode the sample is cooled to 5K before the measuring field(H) is switched on and the measurement is made while warming up the sample. In FCC mode the applied H is switched on in the T regime above the FM-AFM transition temperature (T$_N$) and the measurement is made while cooling across T$_N$. After going down to 5K in the FCC mode, the data points are taken again in the presence of same H while warming up the sample. This is called FCW mode.

\begin{figure}[t]
\centering
\includegraphics[width = 8 cm]{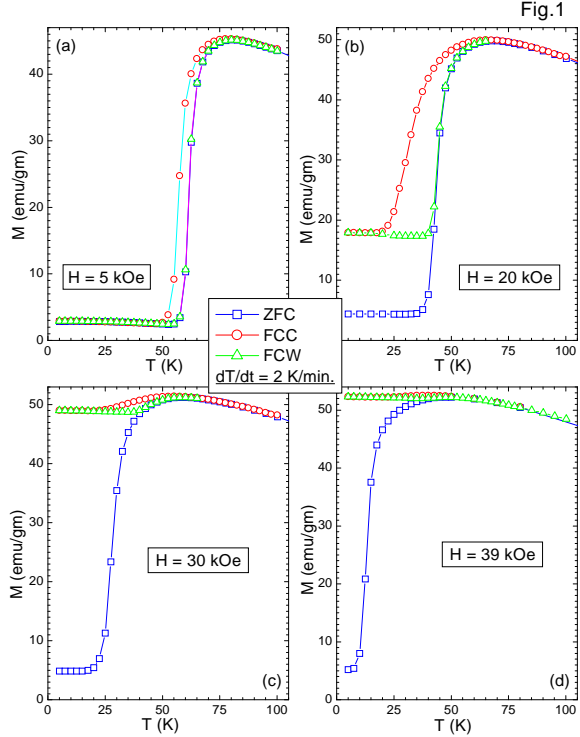}
\caption{Magnetization (M) versus temperature (T) plots for Ce(Fe$_{0.96}$Ru$_{0.04}$)$_2$ sample obtained in zero field cooled (ZFC), field cooled cooling (FCC) and field cooled warming (FCW) mode in applied fields H= 5, 20, 30 and 39 kOe. }
\label{fig1}
\end{figure}

Fig.1 shows the magnetization (M) versus temperature (T) plot of the Ru-doped CeFe$_2$ sample in different applied magnetic fields.   In Fig. 1(a) the FM to AFM transition in the presence of 5 kOe of applied field is marked by the sharp drop in M below 80K and shows substantial thermal hysteresis, which is an essential signature of a first order transition. There is however, no significant difference between M$_{ZFC}$ and M$_{FCW}$ magnetization. The broad nature of the FM to AFM transition is the intrinsic property of the sample and not due to inhomogenity\cite{13,14,16}. This point has been further argued through specific heat measurements\cite{15} on 50 micron size samples (in comparison to few mm size of sample used here) which showed same width and thermal hysteresis of the transition. The T$_N$ is reduced with the increase in H. A striking feature appears in the M-T curve when H is above 15 kOe, namely the M$_{FCC}$(T)  tends to get flattened and does not merge with the M$_{ZFC}$(T) down to the lowest T of measurement. We argue below that this represents the kinetic arrest of  the FM to AFM transition process at the low temperatures giving rise to a non-ergodic magnetically inhomogeneous state with coexisting FM and AFM phases. 

\begin{figure}[t]
\centering
\includegraphics[width = 8 cm]{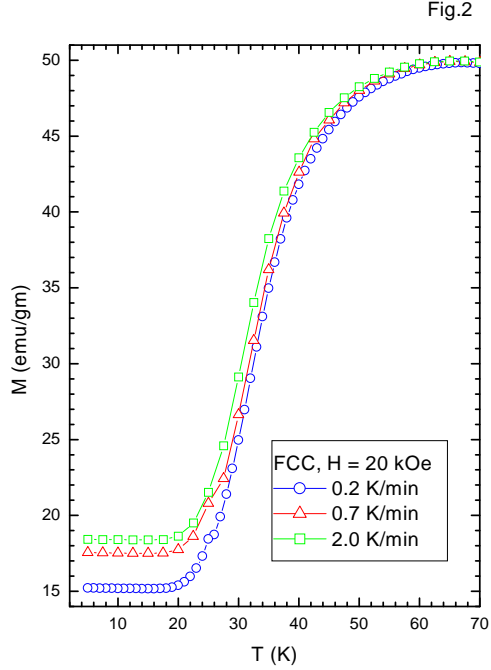}
\caption{Magnetization (M) versus temperature (T) plots for Ce(Fe$_{0.96}$Ru$_{0.04}$)$_2$ sample in applied field of H= 20 kOe obtained in the filed cooled cooling (FCC) mode with three different cooling rates 0.2 K/min, 0.7K/min and 2K/min. }
\label{fig2}
\end{figure}

To elaborate on the supercooling of the FM phase across the FM to AFM transition and the eventual low temperature  kinetic arrest of the transition process, we concentrate on the results of  M-T measurements performed in a field of 20 kOe. An important characteristic of   supercooling and the subsequent glass transformation is the cooling rate dependence. We show in Fig.2 the M-T curve obtained in the FCC mode with cooling rates of  2 K/min, 0.7 K/min and 0.2 K/Min. The effect of cooling rate is particularly striking in the T-region where the kinetic arrest is taking place. M-T curve obtained with the higher cooling rate gets arrested with higher M value which implies higher frozen FM phase fraction. 

We define the onset of glass transformation as the temperature T$_g$ where the M$_{FCC}$(T) starts flattening out without touching M$_{ZFC}$(T). We have earlier defined T$^*$ (the limit of supercooling) as the temperature where the M$_{ZFC}$(T) and M$_{FCC}$(T) merges \cite{14}. When the applied H is below $\approx$15 kOe, M$_{ZFC}$(T) and M$_{FCC}$(T) merges at a finite T and the FM-AFM transition is completed. Accordingly we surmise that T$_g<$T$^*$. Above 15 kOe the FM-AFM transition process is arrested. Here T$_g>$T$^*$ and the low temperature (T$< T_g$) state becomes non-ergodic with FM-AFM phase coexistence. With higher H non-ergodicity sets in at higher temperature, until T$_g$ line meets the T$_N$ line.  More rigorously T$_g$ should be defined as the temperature where the characteristic relaxation time crosses the typical experimental time involved in measurements. This aspect of relaxation effect is discussed below.

We have argued earlier that in the FCC path the FM state persists at temperatures well below T$_N$ as supercooled state\cite{14,16}.This FM state is highly metastable and any energy fluctuations tend to convert it into equilibrium AFM state. To elucidate more on this process we have studied relaxation of  M on various points on the FCC leg of the M-T curve. As the system goes below T$_N$  and approaches the limit of supercooling T* \cite{17},  the barrier height between the metastable (FM) and the equilibrium (AFM) state in the free energy curve decreases and hence a decrease of  M with higher relaxation rate is expected. This is  actually observed in the T-regime 25-40 K. However, some marked change in relaxation takes place in the T-regime below roughly 23 K. The relaxation rate decreases drastically and below 15K the relaxation of M is very small even though the metastable FM state persists. Relaxation data below 23K can be fitted well with Kohlrausch-Williams-Watt stretched exponential function  $\Phi$(t)$\propto$exp[-(t/$\tau)\beta$] (see Fig.3), where $\tau$ is characteristic relaxation time and $\beta$ is a shape parameter between 0.6 and 1. The characteristic relaxation time $\tau$ tends to diverge below 15K (see inset of Fig.3). This behaviour is typical of  what has been observed in many glass-formers in the T-regime of glass formation\cite{1,4}. The non-Arrhenius behaviour of  $\tau$(T) resembles that of fragile-glass former like o-terphenyl\cite{1}.

\begin{figure}[t]
\centering
\includegraphics[width = 8 cm]{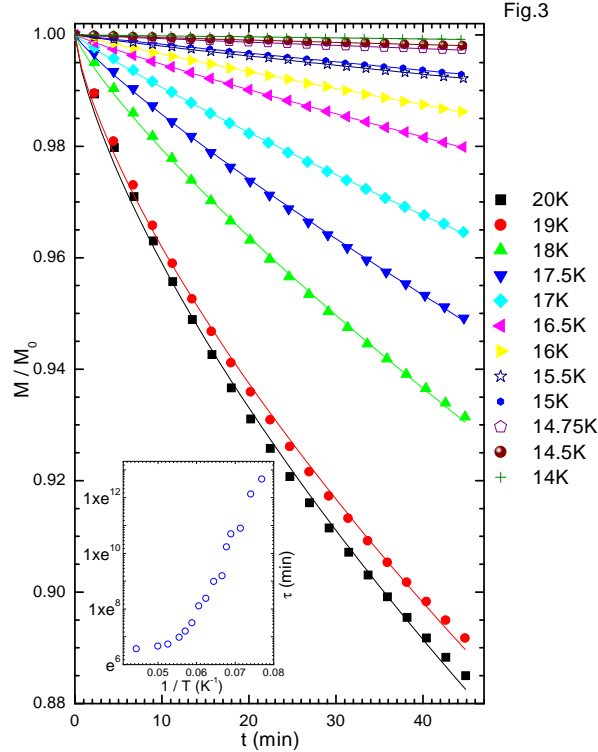}
\caption{Normalized magnetization(M) vs time (t) plots for Ce(Fe$_{0.96}$Ru$_{0.04}$)$_2$ sample at various fixed temperatures on the FCC path. M is normalized with respect to the initial M$_0$ obtained after one 1 sec of stabilizing at the respective temperatures. The applied field is 20 kOe. Each T of measurement is reached with a cooling rate of 2K/min from a starting T=120K which is well within the FM regime. The solid lines represent the fitting with Kohlrausch-Williams-Watt (KWW) exponential function. The inset shows the characteristic relaxation time $\tau$ (obtained from the fitting of KWW function) as a function of T.}
\label{fig3}
\end{figure}

We have earlier shown through micro-Hall probe imaging the coexistence of FM and AFM patches across the transition regime\cite{16}. We argue now that below T$_g$ the supercooled FM patches are frozen in time and they hardly respond to any energy fluctuations. We have found that at various fixed T between 5 and 15K an energy fluctuation produced by a small cycling of temperature hardly caused any change in M. In Fig.4. Curve A1 to A2, drawn with black open squares,  represents the M$_{FCC}$(T)obtained a magnetic field of 20 kOe. On completion  of this initial FCC leg the sample is warmed from 5K to 15K and then cooled back to 5K. Magnetization followed the initial FCC curve and is reversible (data not shown here for the sake of clarity). This highlights the highly viscous nature of the phase configuration. The sample is then warmed up from A2(5K) to A3(20K), then cooled down again to A4(5K) and warmed up to A5(25K). The cycling is followed continuously from A5  to A10 through A6(5K), A7(30K), A8(5K), A9(35K), and A10(5K). With every T-cycle M is reduced. While warming in this FCW mode the non-ergodic FM patches are released from their frozen state. These released FM patches are  susceptible to energy fluctuations and they get converted to the equilibrium AFM state causing a decrease in M. But as the sample is warmed up from A10 to B1(40K), M increases after an initial decrease. In the subsequent cycle B1 to B2(5K), M  shows a trend different from the previous cycles i.e. M registers an initial rise before falling with decreasing T. This initial rise is due to an AFM (superheated) to FM (equilibrium) conversion. Some patches of the AFM state will be retained as the superheated state until the limit of superheating T** \cite{17}  is reached. In this T regime any temperature cycling will cause the conversion of the metastable AFM state to equilibrium FM state and there will be an increase in M. This trend is observed in all the temperature cycling that follow, i.e. B2 to B3 to C1(45K), and C1-to-C2-C3-D1(50K). Cooling below D1, through D2, nearly reproduces the original FCC magnetization curve A1-A2. 

\begin{figure}[t]
\centering
\includegraphics[width = 8cm]{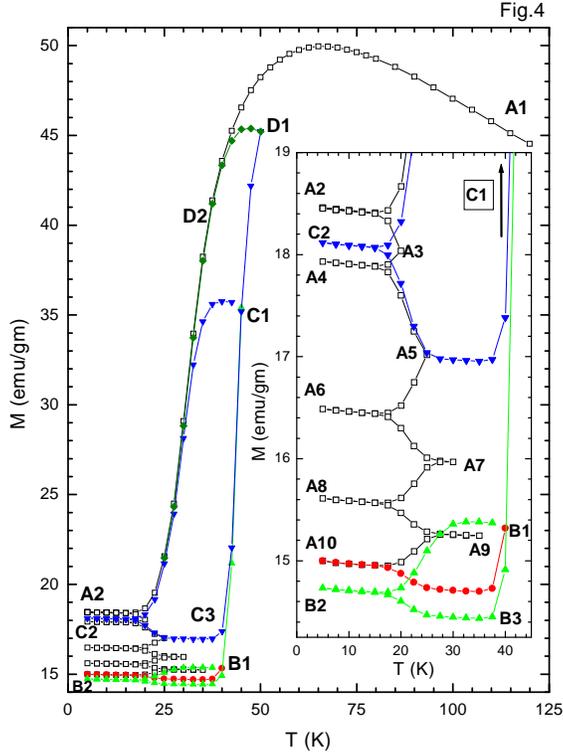}
\caption{Effect of energy fluctuations obtained through temperature cycling at various points on the FCC path. The main panel shows the complete experimental cycle, while the inset depicts a magnified version of the data recorded at low temperatures.  It is to be noted that the shape of the curves below 20K remains identical in all the temperature cycles; they represent magnetic states of the sample with various arrested FM fraction. }
\label{fig4}
\end{figure}

All these results of  M-T measurements on Ce(Fe$_{0.96}$Ru$_{0.04}$)$_2$ alloy with 20 kOe applied field present a striking similarity with the glass formation process in polymeric glass ploy vinyl chloride (widely known as PVC) \cite{18}.In PVC  the glassy state is reached via a heterogeneous state consisting of microcrystallites\cite{19}. In comparison we have here a frozen configuration consisting of metastable FM patches (read liquid ) and AFM patches (read crystallite). Fig.1 indicates that the ferromagnetic fraction in this configuration increases markedly with the increase in applied H and it is  possible to arrest the FM-AFM transition just below T$_N$ at H=39 kOe. Although thermomagnetic irreversibility and metastability are also properties of spin-glasses \cite{6}, the properties observed here are distinctly different\cite{14,16}.  Neutron scattering studies\cite{10} have earlier ruled out unequivocally the existence of any such spin-glass like magnetic ordering in doped-CeFe$_2$ alloys. Instead in the FM to AFM transition region we are dealing with a state with distinct FM-AFM phase co-existence at least in the mesoscopic scale\cite{16}. At low temperatures this FM-AFM phase co-existence develop non-ergodicity and the state can be called a 'magnetic-glass'. Fig.1 clearly indicates that this 'magnetic-glass' state is absent in zero and low magnetic fields. It can only be realized in the presence of  applied fields above a certain critical value. It is an interesting question now, can a liquid (say Helium) which does not reach a glassy state in ambient pressure, be transformed into a glass in the presence of a higher applied pressure? However, quenching a liquid with massive pressure apparatus is not an easy task\cite{7}. In this sense the role of a second thermodynamic variable in the kinetic arrest of a first order transition will be more easily studied using the 'magnetic glass' as a medium. There are already some indications that a kinetic arrest in the doped-CeFe$_2$ alloys can also be achieved in the isothermal H-variation experiments\cite{12}.

In conclusion we have shown explicitly the formation of a new type of 'magnetic-glass' state in a test-bed magnetic system  Ce(Fe$_{0.96}$Ru$_{0.04}$)$_2$ alloy. This glassy state arises from the kinetic arrest of a first order FM to AFM phase transition. The influence of a second thermodynamic variable namely magnetic field is clearly shown in this temperature induced phenomenon. We believe that various manganese oxide compounds showing first order FM-AFM transition and phase separation beaviour\cite{20} can be possible sources for such 'magnetic-glass'.  Some such indications in this regard already exist\cite{21}.  Such magnetic mediums will provide excellent context to study the supercooling and possibility of the onset glassy behaviour in a two parameter (T and H)  phase space, specially where the first order transition temperatures T$_N$ varies strongly with applied magnetic field. This ability to easily vary T$_N$ allows the study of the regimes where the glsss transformation temperature T$_g$ is less than the limit of supercooling T* (as in liquid metals) as well as regimes where T$_g \leq$ T$_N$ (as in standard glass formers like o-terphenyl and PVC) within the same system using a relatively simple experimental setup.

\end{document}